\documentclass[aps,prl,showpacs,twocolumn,preprintnumbers,amsmath,amssymb]{revtex4}
\usepackage{graphicx} 
\usepackage{dcolumn} 
\usepackage{bm} 
\begin{document}
\preprint{Preprint, cond-mat/0509740, revised version, submitted to PRL}
\title{On the stability and growth of single myelin figures}
\author{Ling-Nan Zou}\email{zou@uchicago.edu}
\author{Sidney R. Nagel}
\affiliation{The James Franck Institute and Department of Physics, The University of Chicago, Chicago, IL 60637}
\date{September 26, revised December 17, 2005}

\begin{abstract}
Myelin figures are long thin cylindrical structures that typically grow as a dense tangle when water is added to the concentrated lamellar phase of certain surfactants. We show that, starting from a well-ordered initial state, single myelin figures can be produced in isolation thus allowing a detailed study of their growth and stability.  These structures grow with their base at the exposed edges of bilayer stacks from which material is transported into the myelin.  Myelins only form and grow in the presence of a driving stress; when the stress is removed, the myelins retract.
\end{abstract}
\maketitle

\begin{figure}
\includegraphics[width=3.3in]{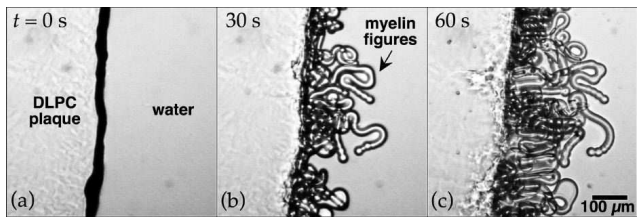}
\caption{The formation of myelin figures from a DLPC plaque sandwiched between two glass slides: (a) just before water contact, and (b) 30 s and (c) 60 s after contact.}
\label{contact}
\end{figure}

	When Dr. Rudolf Virchow looked into a microscope in 1854, he thought he had created living tissue.  What he had discovered instead was what are now called myelin figures  ---  cylindrical structures that grow sinuously and rapidly suggesting a writhing organism \cite{Virchow1854}.  These cylinders readily form when water is added to a dense dry mass of certain surfactants, such as the long-chain phospholipid di-lauroyl phosphaditylcholine (DLPC).  Fig.\ \ref{contact}a-c shows a series of three photographs, taken 30 s apart, of densely packed myelin figures where the cylinders have grown from zero to approximately 100 $\mu\textrm{m}$ in length.  In the concentrated phase, DLPC has a multilamellar structure consisting of stacked planar bilayers.  When water permeates into the stack, the bilayers curl up and form concentric cylinders --- the myelin figures \cite{SakuraiSuzukiSakurai1989}.  How and why do the planar bilayers deform into cylinders and once formed, what drives their growth?

	Previous studies observed that the length $L$ of densely packed myelins grow in time $t$ as $L\propto t^{1/2}$, leading to the suggestion that growth is governed by the collective diffusion of surfactant in solution \cite{SakuraiKawamura1984, Haran&c2002, Taribagil&c2005}. Structurally, Buchanan \textit{et al}. hypothesized myelins develop from a blister-like instability on the planar bilayers and grow in length as the dry surfactant absorbs water and swells in volume \cite{BuchananEgelhaafCates1999}. Finally, as myelins can maintain their cylindrical morphology for several hours, they were considered to be semi-stable structures in local-equilibrium \cite{SakuraiKawamura1984, BuchananEgelhaafCates1999}. These hypotheses have been difficult to validate in experiment because the densely packed myelins formed in these earlier experiments, as shown in Fig.\ \ref{contact}, make it impossible to examine individual myelins in detail.  Here, we show how to grow single myelins from a well-ordered initial material.  Our observations of isolated myelins leads us to question all of these hypotheses.

\begin{figure}
\includegraphics[width=3.3in]{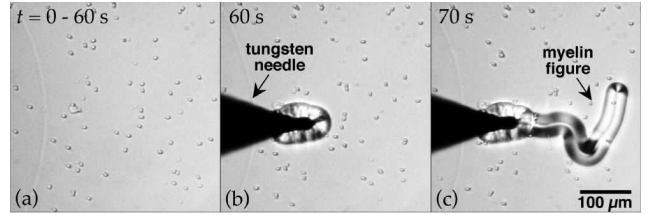}
\caption{The planar surface of a well-ordered DLPC stack shortly after immersion in water: no myelin figures had formed after 60 s (a). A tungsten needle punctures the stack surface (b), causing a myelin figure to form and grow (c).}
\label{immersion}
\end{figure}

	Our method for producing single myelins sheds light upon their structural origins.  We prepare the initial material by depositing a small plaque of DLPC on a glass slide, and incubating it in a humid environment at $60^\circ\textrm{C}$ for 2 days.  This anneals the material so that the bilayers form stacks ordered over large  ($>1 \ \textrm{mm}^2$) regions, as we can check via optical birefringence. We immerse this annealed material in water.  If myelins form via blistering, then myelins should develop on the bilayer stack's planar top surface.  However, we never observe this even after immersion for several hours, as shown in Fig.\ \ref{immersion}a.  However, if we repeat the experiment but, immediately after immersion, puncture the top surface of the stack with a sharp needle, we find myelin growth from the damaged region as shown in Fig.\ \ref{immersion}b-c.  This demonstrates that myelins do not form via blistering of the planar bilayer surface but instead grow from multilamellar edges.  

\begin{figure}
\includegraphics[width=3.3in]{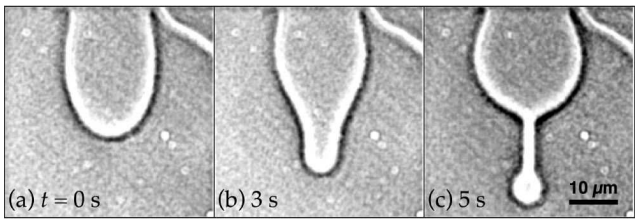}
\caption{The formation of myelins as seen in the drying drop experiment. An initially rounded bilayer stack (a), deforms into an elongated shape (b), and forms a myelin (c).}
\label{drop}
\end{figure}

	We see similar behavior in another experiment.  We take a small drop of a suspension, containing di-myristol phosphaditylcholine (DMPC, a phospholipid similar to DLPC) vesicles dispersed in water, and let it slowly evaporate upon a glass slide.  The vesicles are left as a ring-shaped stain at the perimeter of the drying drop \cite{Deegan&c1997}.   Although most of the stain is disorganized, small well-ordered bilayer stacks (as revealed by optical birefringence) frequently form along the drop contact line.  When such a stack intrudes back into the drop, as in Fig.\ \ref{drop}, it will often develop one or more myelins.  Here a $10 \ \mu\textrm{m}$ long myelin forms from a bilayer stack within a few seconds.  Again, the myelin develops at multilamellar edges and not from blistering on the planar top surface. 

\begin{figure}
\includegraphics[width=3.3in]{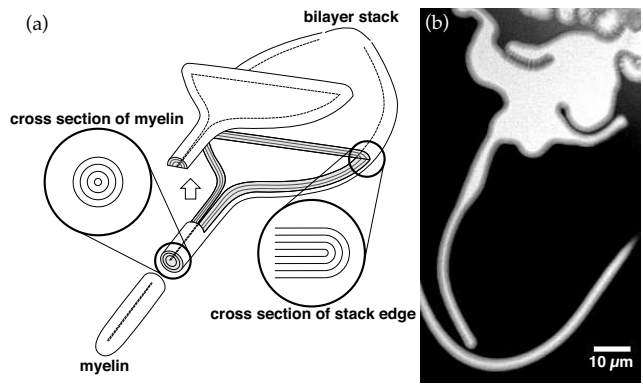}
\caption{(a) Schematic of the myelin/bilayer stack structure of Fig.\ \ref{drop}. Cut-away shows the internal structure at the myelin/bilayer stack junction. Enlargements show the lamellar arrangement at the edge and in the myelin in cross section. (b) Confocal image of a myelin/bilayer stack structure, showing the region of reduced fluorescence along the edge. The laser excitation $\mathbf{k}$ is perpendicular to the image plane.}
\label{structure}
\end{figure}

	These findings suggest that the myelin forms by a protrusion of the stack's rounded edge into the surrounding water. A schematic of this structure is drawn in Fig.\ \ref{structure}a.  This picture give a simple explanation for what sets the diameter of a myelin: it is the same as the thickness of its parent, stacked-bilayer, structure.  Thus the wide distribution of myelin diameters seen in earlier experiments can be attributed to the fact the initial material contained many lamellar domains of various sizes.

	We find evidence for the structure shown in Fig.\ \ref{structure}a by using confocal microscopy to image the junction between the bilayer stack and the myelin. We label the DMPC with 5\% molar fraction of a fluorescent phospholipid, 14:0/12:0-NBD phosphaditylcholine (NBD-PC), whose fluorescence intensity depends on its orientation with respect to the incident laser beam.  This allows us to infer the local lamellar orientation from the fluorescence image because regions where the bilayers are perpendicular to the excitation beam are brighter than where the bilayers are parallel to it. This is particularly clear when the image slice is taken half way between the top and bottom surfaces, as shown in Fig.\ \ref{structure}b.  The fluorescence is reduced in a well-delineated, continuous region of uniform width along the edge of both the bilayer stack and the myelin.  This is where the bilayers are folded over and are locally oriented parallel to the laser as indicated in Fig.\ \ref{structure}a.  Thus the stack thickness, as well as the diameter of the myelin, should be twice the width of the reduced intensity region as we verified, to within $0.3 \ \mu$m, by confocal imaging.

\begin{figure}
\includegraphics[width=2.5in]{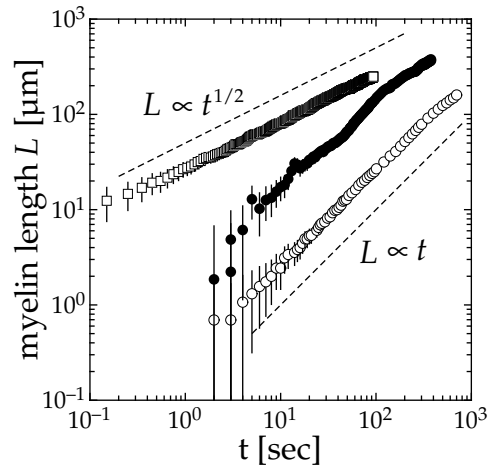}
\caption{Myelin length vs. time: ($\Box$) densely packed myelins formed on direct contact between water and dry surfactant as in Fig.\ \ref{contact}; ($\circ$) single myelin drying drop experiment as in Fig.\ \ref{drop};  ($\bullet$) single myelin grown in immersion-and-puncture experiments as in Fig.\ \ref{immersion}. Dashed lines are  $L\propto t^{1/2}$ and  $L\propto t$ power laws as indicated.}
\label{growth}
\end{figure}

	With the production of single myelins, we can study myelin growth and stability with greater detail than was possible before.  Previous reports found that dense myelin thickets grow as $L \propto t^{1/2}$, reproduced by us in Fig.\ \ref{growth}.  However, we find single myelin growth can have various time-dependences that depend on the experimental geometry, as shown in Fig.\ \ref{growth}.  In the drop experiment of Fig.\ \ref{drop}, growth follows va linear time dependence: $L \propto t$. In the immersion experiment of Fig.\ \ref{immersion}, the growth can also be roughly described by $L \propto t^p$, $p\sim 1$, but the fit is comparatively poor and the power law description may not be correct.  These different growth laws are not easily explained by either the diffusive growth model \cite{SakuraiKawamura1984} or the growth-by-swelling mechanism \cite{BuchananEgelhaafCates1999}.

\begin{figure}
\includegraphics[width=3.0in]{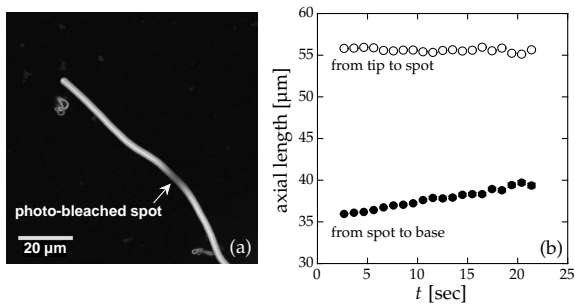}
\caption{Tracking surfactant transport in myelin growth. (a) Confocal microscope image of a fluorescent myelin figure with a photo-bleached dark spot. (b) The position of the spot versus time as measured from the myelin base and from the myelin tip.  The tip-to-spot distance remains constant indicating that growth occurs only at the myelin base.}
\label{spot}
\end{figure}

	Using fluorescence labeling we show a myelin grows via the transport of lipid molecules from the parent structure into the protruding cylinder. We photo-bleach a small section of a growing myelin (labeled with NBD-PC), as shown in Fig.\ \ref{spot}a. This spot broadens, consistent with the known bilayer lateral-diffusion constants \cite{DietrichMerkelTampe1997, FilipovOraddLindblom2003b}.  When we plot in Fig.\ \ref{spot}b the distance, measured along the structure, of the spot's intensity minimum both from the base of the myelin and from its tip, we find the spot translates in parallel with the growing tip.  This suggests the myelin grows via a plug-like influx of lipid through its base. We also did not observe any surfactant concentration gradients (as indicated by the uniform fluorescence prior to photo-bleaching) along the myelinÕs length, in contrast to earlier claims \cite{Sakurai1985}. Both of these observations cast doubt on the diffusive growth model for myelins.  Although they are more consonant with the growth-by-swelling mechanism of Buchanan \textit{et al}., other difficulties remain.  In particular, in the drop experiment, myelins grow steadily even though the drop contact-line is pinned and there is no net flux of water into the dehydrated surfactant on the ring-stain; while in the immersion experiment, the surfactant swells, yet without exposed edges myelins do not form and grow.  In neither experiments does myelin growth (or the lack thereof) appear to be simply related to the swelling of dehydrated surfactant on the influx of water.  A future publication \cite{Zou2005} will describe a growth model via differential hydration, which can successfully reproduce both the $L \propto t$ and $L \propto t^{1/2}$ dependences. 

	We now address the question whether myelins are semi-stable structures in local equilibrium or whether they are dynamic structures developed in response to some driving stress. Our experiments suggest that they are the latter.  As shown earlier, when we create a defect in the top surface of a large well-ordered stack shortly after immersing it in water, myelins grow within a few seconds.  However, if we wait an additional few hours, they stop growing and \textit{retract} into the parent structure.  We suggest retraction occurs once water saturates the bilayer stack, removing the hydration gradient and relaxing its internal stress. We can estimate the saturation time $\tau$ for a stack of $N$ bilayers as $\tau=N^2{d_0}^2/D_\textrm{eff}$, where $D_\textrm{eff}$ is the effective diffusion constant for water permeating through the bilayer stack, and $d_0$ is the bilayer repeat spacing \cite{Tanner1978, DudkoBerezhkovskiiWeiss2004, CarruthersMelchior1983}. For typical experiments, $N\approx1000$, $d_0 = 6$ nm, and $D_\textrm{eff} \approx 6 \times 10^{-3} \ \mu\textrm{m}^2/\textrm{s}$, the saturation time $\tau \approx 1$ hour. This estimated time for stress relaxation is consistent with the observed life-time of these myelins. Moreover, if we repeat this experiment from the beginning, but wait an hour \textit{before} scratching the top surface, myelins do not grow from that defect at all.   Similar behavior is found in the drop experiment of Fig.\ \ref{drop}.  When we halt evaporation, myelin formation quickly ceases and the myelins already formed retract at a rate comparable to their initial growth rate.  These experiments show that without the stress created by a hydration gradient across the surfactant/water interface, myelins not only do not grow but cannot persist.  
	
\begin{figure}
\includegraphics[width=3.3in]{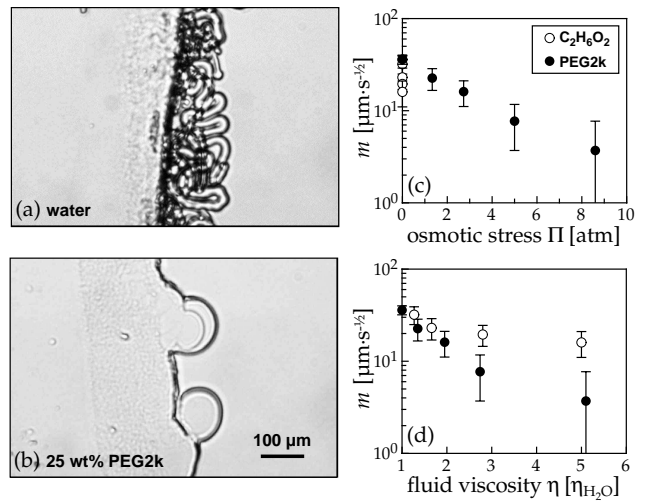}
\caption{A similar experiment as Fig.\ \ref{contact}, showing the myelin growth 10 s after contact with (a) water and (b) 25 wt\% PEG 2000 ($\Pi =12.6$ atm). In the latter case myelin formation was completely suppressed. The growth parameter, $m$, plotted versus (c) imposed osmotic stress $\Pi$, and (d) fluid viscosity $\eta$.  The data are of PEG ($\bullet$) and ethylene glycol ($\circ$) solutions.}
\label{osmotic}
\end{figure}	
	
	We can also suppress myelin growth by lowering the hydration gradient in another manner.  We sandwich dry DLPC between two glass slides, and inject fluid into the gap to make contact with the sample.  This geometry is known to produce myelins that grow as $L = mt^{1/2}$ \cite{BuchananEgelhaafCates1999, Haran&c2002}.  When the fluid is pure water, myelins grow rapidly.  However, when the fluid is an aqueous solution of  polyethylene glycol (PEG 2000, m.w. 2000 Da), growth is slowed and can be suppressed entirely for high enough PEG concentrations.  This is shown in Fig.\ \ref{osmotic}a-b.  PEG 2000, with a radius of gyration $R_g=1.8$ nm, is sterically excluded from stacked DLPC bilayers (maximum bilayer-bilayer separation $=2.7$ nm) and therefore competes for water with DLPC.  This imposes an osmotic stress on the DLPC \cite{ParsegianRandRau2000} acting in opposition to the hydration gradient driving myelin formation. Fig.\ \ref{osmotic}c plots the growth coefficient, $m$, versus the applied osmotic stress, $\Pi$ \footnote{We used the osmotic pressure data collected by R.\ P.\ Rand (\texttt{http://aqueous.labs.brocku.ca/osfile.html}).}.  In order to check whether it is $\Pi$ or the fluid viscosity, $\eta$, that controls the growth in this regime, we performed control experiments using a series of ethylene glycol/water mixtures, each chosen to have viscosity close to one of the PEG solutions.  Since ethylene glycol is too small to be excluded from the bilayer stack, the ethylene glycol solutions should not impose any osmotic stress on the DLPC.  Fig.\ \ref{osmotic}d, plotting the growth coefficient $m$ versus $\eta$, shows that the PEG solutions suppress myelin growth much more dramatically than does the ethylene glycol mixtures for viscosities up to $\eta = 5 \eta_{\textrm{H}_{2}\textrm{O}}$.  And as ethylene glycol is also the PEG monomer, this experiment rules out specific chemical reaction between the lipid and the PEG as the cause of myelin suppression.  These results again illustrates the importance of a driving stress in myelin formation and growth.  In particular, it is difficult for the diffusive growth model to explain why myelin growth is strongly suppressed by applied osmotic pressure, but hardly at all by increased fluid viscosity \cite{SakuraiKawamura1984, Haran&c2002}.

\begin{figure}
\includegraphics[width=3.3in]{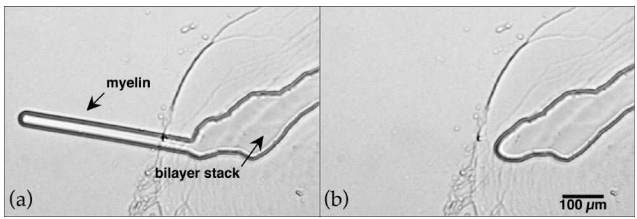}
\caption{The formation of a myelin (a) driven by an external flow, and (b) when the flow is switched off.  In this particular experiment, water flows from right to left at about 1 cm/s.}
\label{flow}
\end{figure}

	A final experiment showing myelins are stress-generated structures uses a flow cell made up of two glass slides with a 0.5 mm gap between them.  Water, controlled by a syringe pump, flows in one end of the cell, which contains well-ordered DLPC stacks, and exits from the other.  The DLPC is first allowed to equilibrate fully in the water for several hours before we turn on the flow.  Without flow, the bilayer stacks are quiescent and no myelins are formed.  However, with a sufficiently strong flow, myelins grow from the edges of the bilayer stacks.  When this flow is stopped, these myelins immediately retract.  This sequence is shown in Fig.\ \ref{flow}. 

	Our experiments show that myelins form via the protrusion of multilamellar edges, rather than blistering on the planar surface.  The myelin diameter is determined by the thickness of its parent structure.  Depending on the experimental geometry, myelin growth exhibits different time-dependences, suggesting the process is not simply governed by the collective diffusion of surfactant in solution \cite{SakuraiKawamura1984} or by the swelling of dry surfactant on the influx water \cite{BuchananEgelhaafCates1999}. Myelins do not develop and cannot persist in the absence a driving stress; when stress is relaxed they retract back into their parent structures. This is contrary to suggestions that myelins are semi-stable and can be explained via free-energy minimization around a local equilibrium, e.g. reference \cite{HuangZouWitten2005}. That myelins can form in response to both an internal thermodynamic stress (hydration gradient) and an external mechanical stress (fluid flow), suggests their formation may instead be due to an elastic instability of stressed bilayer stacks. 
	
	There are other outstanding questions.  If myelin formation is indeed caused by an elastic instability, the elastic properties of myelins and their parent structure should be measured.  While we know the stress from a hydration gradient can drive myelin growth, the actual microscopic processes underlying growth have not been identified and measured.  Finally, myelins are famous for adopting complicated braided and coiled conformations \cite{Haran&c2002, BuchananEgelhaafCates1999}.  What are the forces and conditions that cause such beautiful configurations?  We believe that progress can be made on all these questions now that we have ability to create and study single myelins in isolation.

	We thank  J.-R. Huang, T. A. Witten, M. E. Cates, and L. Xu for their comments.   This work was supported by NSF MRSEC DMR-0213745, NSF DMR-0352777, and the U.S. Department of Education GAANN program.

\end{document}